\documentclass[%
 reprint,
amsmath,stmaryrd,amssymb,textcomp
 aps,
pra,
]{revtex4-1}

\usepackage{graphicx}
\usepackage[justification=centerlast]{caption}
\begin{document}

\title{Controlled transportation of mesoscopic particles by enhanced spin orbit interaction of light in an optical trap}


\author{Basudev Roy$^1$}
\author{Nirmalya Ghosh$^1$} \email{nghosh@iiserkol.ac.in}
\author{S. Dutta Gupta$^2$}
\author{Prasanta K. Panigrahi$^1$}
\author{Soumyajit Roy$^3$}
\author{Ayan Banerjee$^1$} \email{ayan@iiserkol.ac.in}
\affiliation{$^1${Department of Physical Sciences, IISER-Kolkata, Mohanpur 741252, India}}
\affiliation{$^2$School of Physics, University of Hyderabad, Hyderabad 500046, India}
\affiliation{$^3${EFAML, Materials Science Centre, Department of Chemical Sciences, IISER-Kolkata, Mohanpur 741252, India}}
\date{\today}
\begin{abstract}
We study the effects of the spin orbit interaction (SOI) of light in an optical trap and show that the propagation of the tightly focused trapping beam in a stratified medium can lead to significantly enhanced  SOI. For a plane polarized incident beam the SOI manifests itself by giving rise to a strong anisotropic linear diattenuation effect which produces  polarization-dependent off-axis high intensity side lobes near the focal plane of the trap. Single micron-sized asymmetric particles  can be trapped in the side lobes, and transported over circular paths by a rotation of the plane of input polarization. We demonstrate such controlled motion on single pea-pod shaped single soft oxometalate (SOM) particles of dimension around  $1\times 0.5~\mu$m over lengths up to $\sim$15 $\mu$m . The observed effects are supported by calculations of the intensity profiles based on a variation of the Debye-Wolf approach. The enhanced SOI could thus be used as a generic means of transporting mesoscopic asymmetric particles in an optical trap without the use of complex optical beams or changing the alignment of the beam into the trap.

\begin{description}
\item[PACS numbers: 42.50.Tx, 42.25.Ja, 87.80.Cc, 42.50.Wk]
\end{description}

\end{abstract}

\maketitle
\noindent 
Optical spin orbit interaction (SOI) causes an intrinsic coupling between the polarization and spatial trajectory of light and has emerged as a powerful tool to probe light matter interactions. It has led to interesting phenomenon such as the interconversion between spin angular momentum (SAM) and orbital angular momentum (OAM) of photons in isotropic homogeneous media using a tightly focused beam \cite{zhao07} or scattering by mesoscopic particles \cite{haef09, vuo10},
 and the spin Hall effect of light where circularly polarized beams undergo transverse spatial shifts while traversing inhomogeneous refractive index media \cite{ono04, bli08}.  However, while the SOI in anisotropic media is fairly large and can be observed readily in the far-field itself \cite{mar06, berr05}, that in isotropic and homogeneous/inhomogeneous media caused by tight focusing  or scattering is a rather minute effect and is manifested in sub-wavelength scales \cite{zhao07, zeev07, blio08, haef09, vuo10}. Recently, Rodr\'{i}guez-Herrera et al. \cite{her10} have managed to obtain far field signatures of the spin Hall effect as a function of nanoparticle displacements. Such techniques exploiting the various signatures of SOI hold promise as sensitive nano-probes for practical applications.

\begin{figure}[h!t!]
{\includegraphics[scale=0.4]{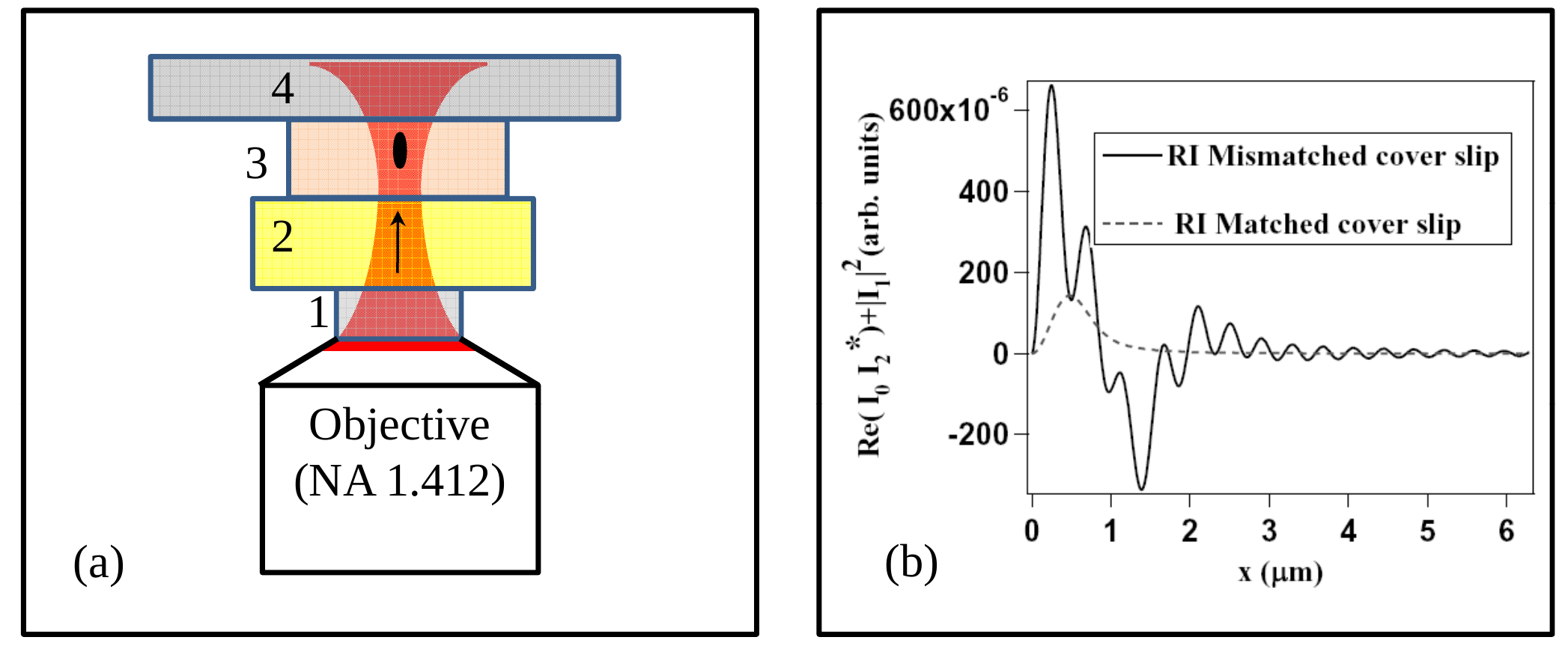}}
\captionsetup{format=plain}
\caption[]{(Color online)(a) Stratified medium (not to scale) formed in our experimental system with a Gaussian beam (direction shown) focused inside the sample solution. The various layers are (1) Objective immersion oil (RI 1.516), (2) Cover slip (RI 1.575), (3) Aqueous solution of micro-particles (RI 1.33), (4) Top glass slide (RI 1.516). For conventional tweezers, the RI of (1) and (2) are matched. (b) Theoretical calculation of the radial variation of $\mathfrak{D}(\rho)$ in the sample chamber for stratified medium with our cover slips that are RI mismatched with objective immersion oil  (solid line),  and those used conventionally (dashed line) with RI (1.516) matched with immersion oil. Both plots are for an axial distance 1 $\mu$m away from the beam focus.}
\label{exptsys}
\end{figure}

\begin{figure}[h!t!]
 \centering{\includegraphics[scale=0.37]{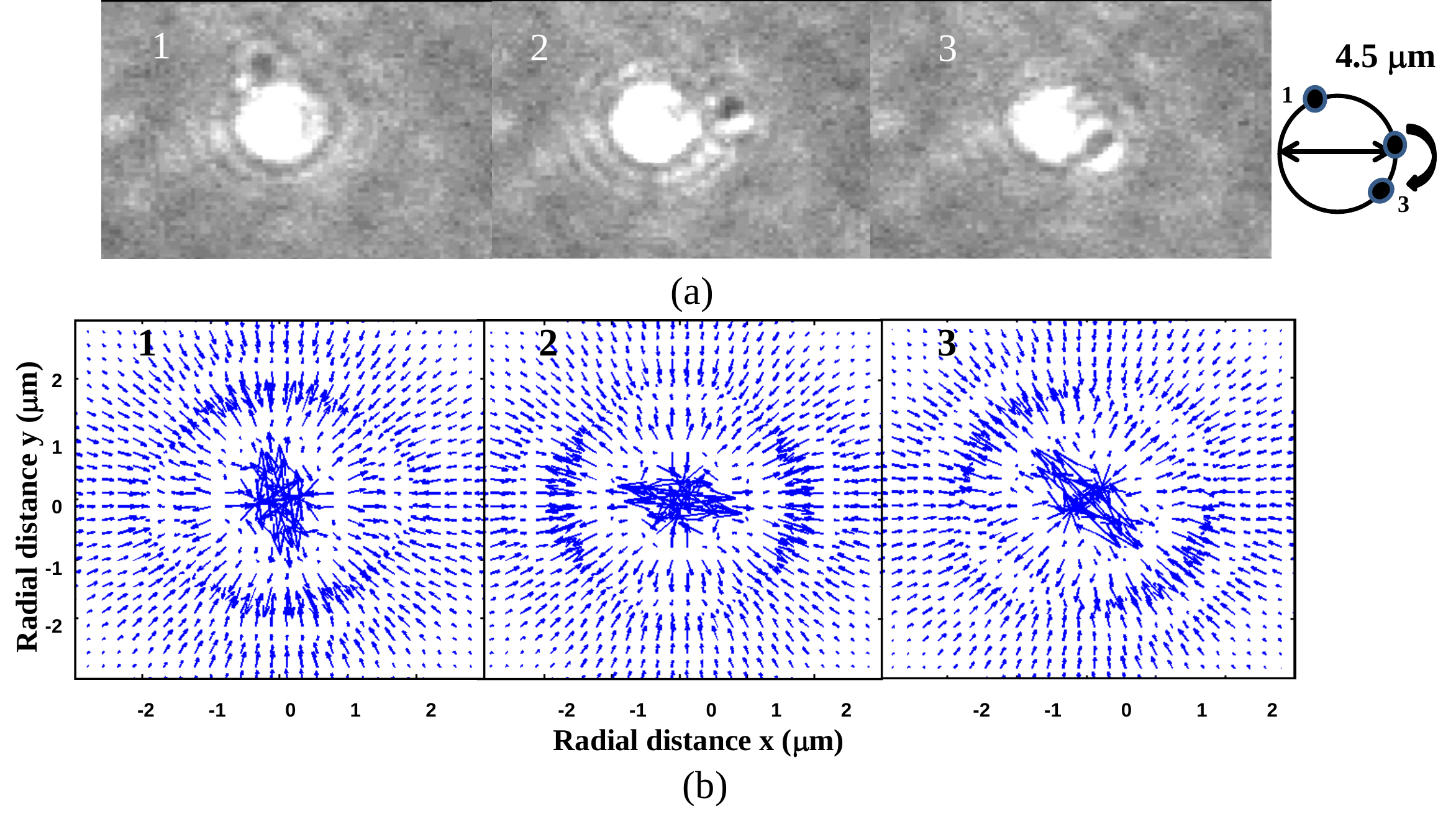}}
    \caption[fieldmap]{(Color online) (a) Time series of three positions of a peapod moved along the periphery of the ring equivalent to a distance $\sim 14~ \mu$m by rotating the linear retarder at the input of the optical trap. (b) shows quiver plots representing theoretical simulations of the intensity gradient at three different polarization angles (approximate) of the input Gaussian beam corresponding to the positions of the trapped peapod in (a). The areas with higher densities of converging arrows indicate local conservative force fields.}
\label{fieldmap}
\end{figure}

In this letter, we experimentally demonstrate controlled transport of mesoscopic particles in a `non-standard' optical trap (tweezers) by exploiting SOI of light.  It is well known that tight focusing of a linearly polarized beam leads to an asymmetry in intensity distribution at the focal plane \cite{roh05} as has been seen in optical traps, which can be interpreted as a manifestation of space-dependent geometric phases (Berry phase) acquired by the focused light \cite{zeev07}. Our setup is distinct from the standard optical tweezers since we use a cover slip that is not refractive index (RI) matched with the immersion oil, leading to an additional dielectric interface (Fig.~\ref{exptsys}a). We demonstrate that this additional interface can lead to well-defined off-axis intensity lobes near the focal plane due to a large anisotropic linear diattenuation (differential attenuation of orthogonal polarizations) effect. These high-intensity lobes occur along the polarization direction on the background of a ring-like intensity profile and possess sufficient field gradient to trap particles, which can then be robustly transported by a change in the input polarization direction  (see Fig.~\ref{fieldmap}).  We achieve trajectory lengths of around 15 $\mu$m by this method without compromising on the trap stiffness since the trapping beam direction remains unaltered.

{\it Experiment and observations:} The experimental setup is a standard optical tweezers configuration consisting of an inverted microscope (Carl Zeiss Axiovert.A1), and a trapping laser (wavelength 1064 nm, TEM$_{00}$) coupled into the microscope back port. The laser beam is tightly focused on the sample using a high numerical oil immersion aperture objective (Zeiss 100X, 1.41 NA) lens. The laser beam is linearly polarized, and the angle of polarization can be controlled by a half-wave retarder placed at the input of the trap. About 25 $\mu$l of the sample (microparticles) in an aqueous solution (1:10000 dilution) is placed in a sample chamber consisting of a microscope glass slide (1 mm thickness) and cover slip, making up the top and bottom surfaces of the chamber respectively. The cover slips we use have a different refractive index (RI) (1.575 at 1064 nm) from the microscope immersion oil (1.516 at 1064 nm), and are also thicker (250 $\mu$m) than the cover slips used conventionally (thickness 130-160 $\mu$m). As shown in Fig.~\ref{exptsys} (a), the entire sample system after the microscope objective thus consists of the following different media: 1) immersion oil, 2) cover slip, 3) sample aqueous solution, 4) glass slide. It is important to note that the forward propagating tightly focused laser beam encounters two RI interfaces (thus forming a stratified medium in the forward direction itself) before it is incident on the top glass side. This is not the case in conventional optical traps where the cover slips are RI-matched with the immersion oil, so that there is a single RI interface for the forward propagating light.

Recently, we demonstrated self assembled closed ring structures of micron-sized polystyrene beads in our trap using such a stratified medium \cite{hal12}. The intensity distribution near the focal plane in this case, as imaged and also theoretically derived in Ref.~\cite{hal12}), resembles a ring structure where beads could be trapped.  In the current study, we make use of the azimuthal asymmetry in this intensity distribution and its polarization dependence caused by enhanced SOI in our system to trap single asymmetric particles  and transport them controllably along the ring. The shape asymmetry of the particles enables them to be preferentially aligned with the polarization direction of the electric field \cite{fri98} near the trap focus so that they can be transported when the direction is changed. The particles we use are single soft oxometalates (SOM) that are synthesized from well-defined molecular precursors (Ammonium phosphomolybdate, $(NH_4)_3[PMo_{12}O_{40}]$)) \cite{sou11}.  The particles have a longitudinal dimension of 1-3 $\mu$m and a lateral dimension of around 500 nm (similar to a shape of a `pea-pod').  In addition, the pea-pods being soft oxometalates, are lower in mass than polystyrene beads of the same diameter (say around 1 - 2 $\mu$m) and can be transported relatively easily by the optical force generated when the input polarization is modified. As shown in Fig.~\ref{fieldmap},  we are able to transport pea-pods over ring diameters between 2 - 5 $\mu$m corresponding to periphery lengths of 6 - 15 $\mu$m by varying the $z$-focus of the trapping microscope \cite{hal12}. A time series plot showing four CCD images taken of a pea-pod trapped in the ring and then translated along the periphery by varying the polarization angle is shown in Fig.~\ref{fieldmap}. The central bright spot and surrounding ring-like intensity structure for the trapping laser is also visible. The maximum diameter of transport we achieve in our experiments is around 5 $\mu$m - beyond which the trap becomes too weak as we discuss later. In what follows, we show that these results can be explained in the context of the enhanced optical SOI in our system.

{\it Theoretical analysis:} In order to study the SOI of a tightly focused fundamental Gaussian laser beam propagating through a stratified medium, we invoke the Debye-Wolf theory \cite{wolf59, rich59, bliop11, her10}, where an incident collimated Gaussian beam is decomposed into a superposition of plane waves having an infinite number of spatial harmonics. After focusing by a high NA lens, the resulting field amplitude can be related to the incident field  by the action of a transfer function that can be written as $A = R_z(-\phi)R_y(\theta)R_z(\phi)$, where $R_i(\alpha),~i=x, y, z$ represents the SO(3) rotation matrix around the $i$ axis by angle $\alpha$. $\phi$ is the azimuthal angle, while $\theta$ is the polar angle defined with respect to $x$ and $z$ axis of the laboratory frame, respectively. Now, in dealing with stratified media one needs to account for the polarization content of each spatial harmonic (details provided in Section I, Supplementary Material A). Then, the resultant field amplitude $\vec{E}_{res}(\theta,\phi)$ can be written in terms of the incident amplitude $\vec{E}_{inc}(\theta,\phi)$ as
\begin{equation}
\label{fieldinout}
\vec{E}_{res}(\theta,\phi)=A\vec{E}_{inc}(\theta,\phi),
\end{equation}
where the transfer function $A$ is given by 
\begin{equation}
\label{transfermatrix}
A = \left[
\begin{array}{lll}
\ a - b\cos 2\phi  &\ -b\sin 2\phi &\ c\cos\phi\ \\
\  -b\sin 2\phi &\ a + b\cos 2\phi &\ c\sin\phi\ \\
\ -c\cos\phi &\ -c\sin\phi &\ a-b\ \\
\end{array}
\right ].
\end{equation}
For the forward-propagating case, the coefficients of $A$ are given by $a = \frac{1}{2} \left (T_s+T_p \cos\theta \right )$, $b = \frac{1}{2}\left (T_s-T_p \cos\theta \right )$, and $c =  T_p\sin\theta$, where $T_s$ and $T_p$ are the amplitude transmission coefficients for the s- and p-polarized components. In general, $\vec{E}_{res}(\theta,\phi)$  is a superposition of forward and backward propagating waves in the stratified medium, though the dominant contribution comes from the forward propagating waves. For the backward propagating waves the coefficients in Eq.~\ref{transfermatrix} would be modified with $\theta$ replaced by $\pi - \theta$, and the Fresnel reflection coefficients $R_s$ and $R_p$ being used instead of the transmission ones. The final field can be obtained by integrating Eq.~\ref{fieldinout} over $\theta$ and $\phi$, so that we have
\begin{eqnarray}\label{polarint}
\vec{E}(\rho,\psi,z)&=&  i\frac{kfe^{-ikf}}{2\pi}\int_0^{\theta_{max}}\int_0^{2\pi}
\vec{E}_{res}(\theta,\phi)e^{ikz\cos\theta}\nonumber\\& \times & e^{ik\rho\sin\theta\cos(\phi-\psi)}
sin(\theta)\>d\theta d\phi,
\end{eqnarray}
where $r$ is set to $f$ -- the focal length of the lens, and the limit for the $\theta$ integral is set by the numerical aperture of the microscope objective.
The cylindrical coordinate system is chosen for the convenience it offers to track the polarization of the light beam at the output of a high numerical aperture objective which forces a drastic modification of the initial polarization \cite{roh05}. For an incident linearly polarized beam of light (polarized along $x$ direction represented by a Jones vector  $\left[\ 1\ 0\ 0\ \right]^T$), the electric field inside the medium can be written from Eqn.~\ref{polarint} in  matrix form as 
\begin{eqnarray}\label{fieldout}
\left[
\begin{array}{c}
 {E_x}\\
 {E_y}\\
 {E_z}\\
\end{array}
\right ] &=& C \left[
\begin{array}{lll}
 I_0 + I_2\cos 2\psi  & I_2\sin 2\psi & 2i I_1\cos\psi \\
 I_2\sin 2\psi & I_0 - I_2\cos 2\psi & 2iI_1\sin\psi \\
 -2iI_1\cos\psi & -2iI_1\sin\psi & I_0+I_2 \\
\end{array}
\right ]\nonumber\\  && \times \left[
\begin{array}{c}
 {1}\\
 {0}\\
 {0}\\
\end{array}
\right ]
 =  C
\left[
\begin{array}{c}
 {I_0+I_2\cos 2\psi }\\
 {I_2\sin 2\psi }\\
 {-i2I_1 \cos \psi }\\
\end{array}
\right ],
\end{eqnarray}
where, the expressions for $I_0(\rho), ~I_1(\rho)$ and $I_2(\rho)$ (note that the $\rho$ dependence of these coefficients is implicitly assumed henceforth) are given in Section I, Supplementary Material A, and $C$ is a constant. We proceed to demonstrate that $I_0,~I_1$, and $I_2$ play a significant role in determining the nature of SOI and the resulting intensity distribution in the focal plane.

A preliminary insight into the nature of the SOI can be gained by considering the topological phase evolution of each of the constituent circular polarization modes of the incident linearly polarized light.  For example, we consider the right circular mode characterized by the three-component cartesian Jones vector   $\left[\ 1~j~0\ \right]^T$. The resulting field obtained from the 3 $\times$ 3 transformation matrix of Eq.~\ref{fieldout} can be represented in terms of three uniform polarization components as follows \cite{zeev07}
\begin{equation}
\label{jonesvec}
E = I_0\left[
\begin{array}{c}
\ {1}\\
\ {i}\\
\ {0}\\
\end{array}
\right ]
+ I_2exp(i2\psi)
\left[
\begin{array}{c}
\ {1}\\
\ {-i}\\
\ {0}\\\end{array}
\right ] -2iI_1exp(i\psi)
\left[
\begin{array}{c}
\ {0}\\
\ {0}\\
\ {1}\\\end{array}
\right ],
\end{equation}
where, for convenience, we have omitted the coefficient $C$ which is common to all the three components. The first component in Eq.~\ref{jonesvec} has the same helicity as that of the incident circular polarization, the second component has opposite helicity and is associated with an orbital angular momentum component of $l =2$, while the third component is linearly polarized carrying orbital angular momentum of $l = 1$. The associated coefficients $I_2(\rho)$ and $I_1(\rho)$ of the transverse (2nd term) and the longitudinal (3rd term) field components thus determine the strength of the spin orbit angular momentum conversion.  

The spatial intensity profile  for incident linearly polarized light --  represented by Jones vectors $\left[\ 1\ 0\ 0\ \right]^T$  and  $\left[\ 0\ 1\ 0\ \right]^T$, respectively --  can then be written using Eq.~\ref{fieldout} as
\begin{equation}
I(\rho) = \ \left|I_0\right|{^2} + \left|I_2\right|{^2}  \pm  2 {\bf Re}(I_0I_2^{\star})\cos 2\psi 
+ 2 \left|I_1\right|^2 (1 \pm \cos 2 \psi).
\label{intensitylinpolar}
\end{equation}
The positive and negative signs are for $x$ and $y$-polarization incident states respectively. It is to be noted that the presence of  ${(\bf Re}\ (I_0I_2^{\star})+ \left|I_1\right|^2)\cos 2\psi = \mathfrak{D}(\rho)\cos 2\psi$ term  in Eq.~\ref{intensitylinpolar} leads to a linear diattenuation effect (diattenuation parameter $\mathfrak{D}$ being well known in conventional polarimetry). Thus, there is a strong  dependence of the spatial intensity distribution both in the radial as well as azimuthal direction on the polarization of the incident beam. Note that $\mathfrak{D}$ would have a non-zero value  even for completely real and non-zero values of $I_0(\rho)$ and $I_2(\rho)$. Thus, the anisotropic diattenuation effect would be present for tight focusing alone (without a stratified medium), which is responsible for the observed elongation of the focal spot along the direction of the incident linear polarization vector \cite{zeev07}. However, as mentioned earlier, the elongation in such situations is a only a sub-wavelength effect. In contrast, for propagation of a tightly focused beam through a stratified medium, the maximum value of $\mathfrak{D}$ (which can be used as a measure of the strength of the SOI) can be both enhanced significantly and shifted off-axis due to multiple reflections at the interfaces of the stratified medium with the chosen refractive index contrasts. The mathematical origin of the enhanced SOI in the system due to the increased values of $I_2$ and $I_1$ (Eq.~\ref{jonesvec}) in a stratified medium.

{\it Results and discussions:} In Fig.~\ref{exptsys}(b), we plot $\mathfrak{D}$ inside our sample chamber as a function of radial distance for input $x$-polarized light for the RI-mismatched cover slips we use (shown in solid lines), and the RI-matched cover slips (dashed lines) used in conventional optical tweezers experiments. The sample thickness is 20 $\mu$m, and the geometrical beam focus is at an axial  distance of 13 $\mu$m inside the chamber. Note that the actual focus gets shifted to around 18 $\mu$m for the RI-mismatched case due to the effect of the stratified medium \cite{hal12}, so that the input beam focuses very close to the top slide.  Also, the off-axis ($x = 1.3~\mu$m) value of $\mathfrak{D}$ for the RI-mismatched cover slips is about 25 times higher than the corresponding value for the RI-matched ones, indicating significantly enhanced SOI. The low off-axis value of $\mathfrak{D}$ for the RI-matched case signifies that the light intensity in the trapping plane is concentrated around the trap axis as shown in Fig.~1 of  Section II, Supplementary Information A. Thus, particles are trapped only in the center - a fact we have verified routinely in our experiments using RI-matched cover slips \cite{hal12}.
\begin{figure}
 \centering{\includegraphics[scale=0.35]{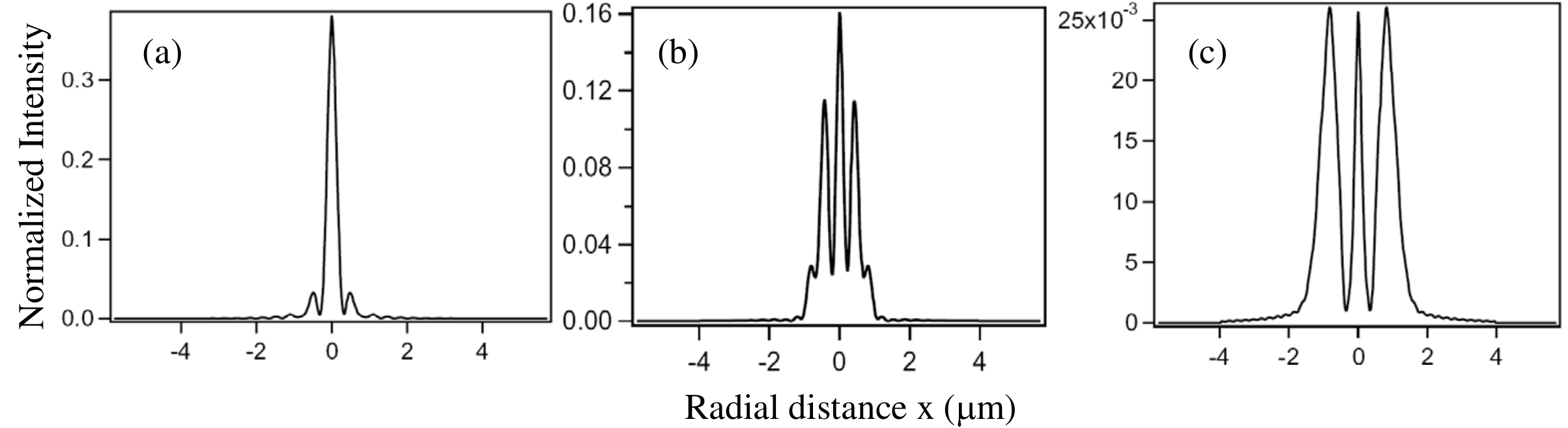}}
\caption[]{(Color online) Simulation of the radial variation of total intensity from Eq.~\ref{intensitylinpolar} for our experimental system at (a) beam focus, and (b) axial distance of 1 $\mu$m away from focus, (c) axial distance of 2 $\mu$m away from focus.}
\label{intensityplot}
\end{figure}

The effect of a high magnitude of $\mathfrak{D}$ on the total intensity is shown in Fig.~\ref{intensityplot}. The figure shows the radial variation of the total intensity as given by Eq.~\ref{intensitylinpolar} at different axial ($z$) distances ($z = 0, 1,~ {\rm and} ~2~ \mu$m, where $z = 0$ corresponds to the focus) inside our sample chamber. Side lobes exist even at the focus (Fig.~\ref{intensityplot}(a)), and these become stronger at $z = 1 ~{\rm and}~ 2 ~ \mu$m (Fig.~\ref{intensityplot}(b) and (c)), with the distance between the lobes in Fig.~\ref{intensityplot}(c) being around 3.5 $\mu$m. It is possible to trap particles in these side lobes as long as the axial trapping is also stabilized by the standing wave cavity formed axially by interference between transmitted and reflected components (the latter from the fourth layer or water/top slide RI interface) of the propagating light.  Stable off-axis trapping at the radial side lobes thus occurs when an axial interference maxima coincides with one of the radial side lobe maximas. Typical axial fringes produced are  shown in Fig.~3 of Section III, Supplementary Information A. It is also observed from Fig.~\ref{intensityplot} that the peak intensity of the side lobes at $z=2~\mu$m is about 14 times less than that at $z=0~\mu$m, and a simple calculation of the intensity gradient reveals that the corresponding trapping force reduces by around 20 times. This sets a limit of around 5 $\mu$m as the maximum distance between side lobes where a particle can still be trapped in a stable manner. The quiver plots in Fig.~\ref{fieldmap}(b) demonstrate that the radial intensity gradient profile resembles an annular ring with two local regions of conservative force fields (represented by a high density of converging arrows) caused by the anisotropic linear diattenuation arising due to enhanced SOI. A particle would thus be preferentially trapped in such a force field and could be transported along the ring by changing the input polarization angle of the trapping beam. This is shown in Fig.~\ref{fieldmap}(b), where we perform simulations of the total field gradient profile explaining our experimental results with sub-plots (a), (b), and (c) signifying input polarization angles ($\psi$) approximately similar to that corresponding to the positions of the trapped peapod in Fig.~\ref{fieldmap}(a). Also, it follows that a change in the optical thickness of the constituent layers in the stratified medium would lead to a change in the radial intensity distribution resulting in different transportation distances for particles as explained in Section IV, Supplementary Information A


This method of particle transport has an dditional advantage of ensuring a constant trap stiffness (since the trapping beam is not moved) in contrast to techniques where particles are transported by scanning the trapping beam itself. Also, while motion of particles in  ring-like trajectories can be achieved using Laguerre-Gaussian beams \cite{mil11} or by holographic tweezers  \cite{gri06}, these methods have a disadvantage wherein any attempt to localize the particle at a particular spatial position would require switching the beam off, in which case the particle would no longer be trapped. 

In conclusion, we have demonstrated enhanced SOI in a tightly focused Gaussian beam propagating through a stratified medium representing an optical trap. The stratified medium is produced by using cover slips of RI different from that of the microscope objective immersion oil. The enhanced SOI manifests itself in an anisotropic linear diattenuation effect which we utilize to controllably transport pea-pod shaped mesoscopic particles over circular trajectories of length around 15 $\mu$m. The SOI could also lead to other interesting consequences such as  space-dependent linear retardance properties that are being presently investigated by us.  

The authors would like to acknowledge Arijit Haldar for the EM field analysis, and Atharva Sahasrabuddhe and Bibudha Parasar for help in preparing the pea-pod samples. This work was supported by the Indian Institute of Science Education and Research, Kolkata, an autonomous research and teaching institute funded by the Ministry of Human Resource Development, Govt. of India.

\nocite{*}


\providecommand{\noopsort}[1]{}\providecommand{\singleletter}[1]{#1}
%




\end{document}


\title{Supplementary Information A}
\author{Basudev Roy$^1$}
\author{Nirmalya Ghosh$^1$} \email{nghosh@iiserkol.ac.in}
\author{S. Dutta Gupta$^2$}
\author{Prasanta K. Panigrahi$^1$}
\author{Soumyajit Roy$^3$}
\author{Ayan Banerjee$^1$} \email{ayan@iiserkol.ac.in}
\affiliation{$^1${Department of Physical Sciences, IISER-Kolkata, Mohanpur 741252, India}}
\affiliation{$^2$School of Physics, University of Hyderabad, Hyderabad 500046, India}
\affiliation{$^3${EFAML, Materials Science Centre, Department of Chemical Sciences, IISER-Kolkata, Mohanpur 741252, India}}
\maketitle
\noindent 
\section{\label{emfield}Theoretical representation of electric field at the focal plane due to tight focusing of polarized light propagating through stratified media}
\begin{figure}[]
\centering{\includegraphics[scale=0.4]{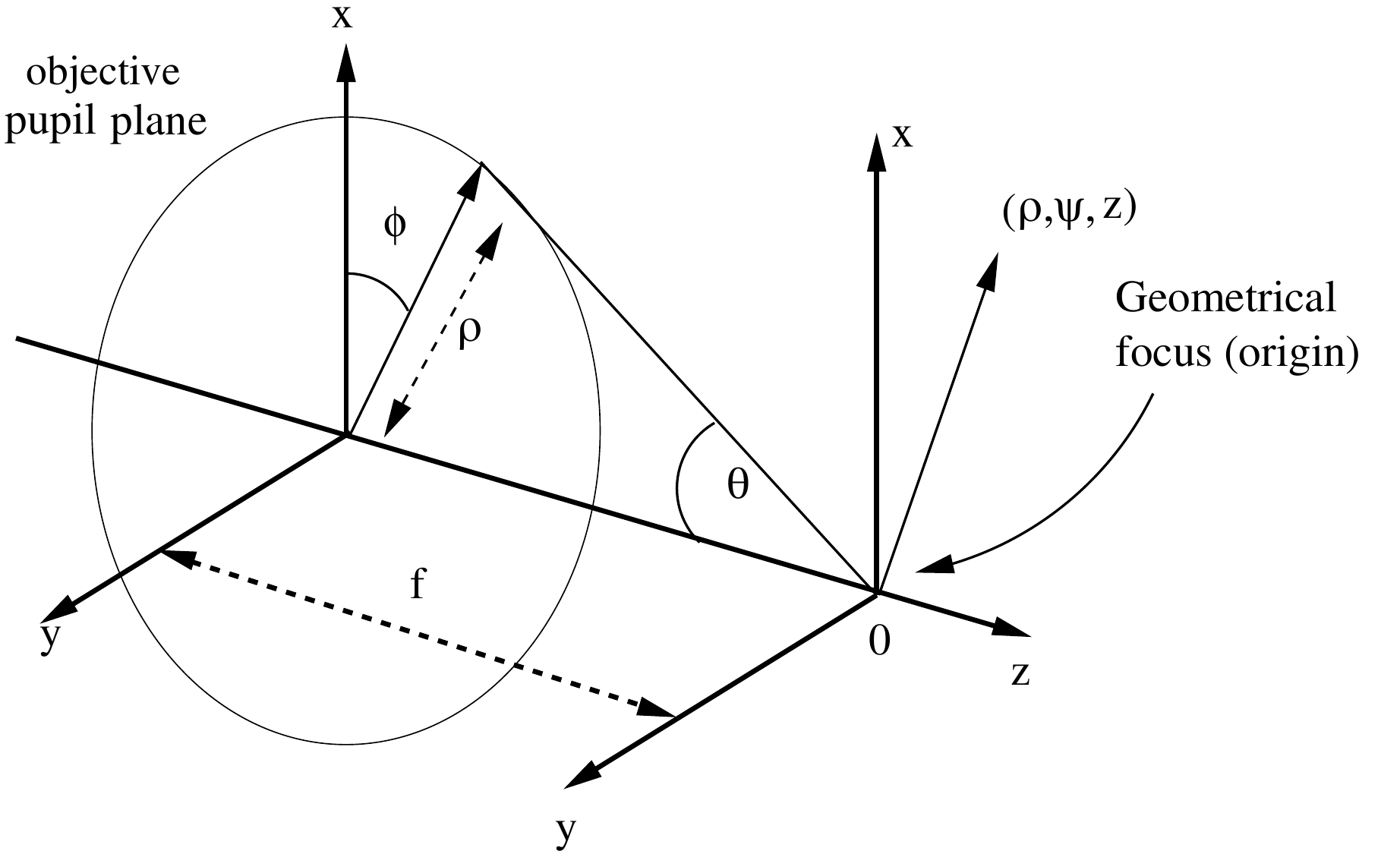}}
\caption[i0i2comp]{(Color online) Coordinate system used in the field analysis.}
\label{coord}
\end{figure}

As per the Debye-Wolf theory \cite{wolf59, rich59, bliop11}, an incident collimated Gaussian beam is decomposed into a superposition of plane waves having an infinite number of spatial harmonics. After focusing by a high NA lens, the resulting field amplitude can be related to the incident field  by the action of a transfer function that can be written as $A = R_z(-\phi)R_y(\theta)R_z(\phi)$, where $R_i(\alpha),~i=x, y, z$ represents the SO(3) rotation matrix around the $i$ axis by angle $\alpha$. $\phi$ could be understood as the azimuthal angle, while $\theta$ is the polar angle defined with respect to $x$ and $z$ axis of the laboratory frame respectively. The coordinate system we use is shown in Fig.~\ref{coord}. 
For focusing into stratified media one needs to take into account the polarization dependence of the field propagating in the media. Thus $A$ needs to incorporate  $T_s~(R_s)$ and $T_p~(R_p)$ -- the Fresnel transmission (reflection) coefficients (generally complex) which include the multiple interface contributions for $s$ and $p$ polarizations respectively. Then, the resultant field amplitude $\vec{E}_{res}(\theta,\phi)$ can be written in terms of the incident amplitude $\vec{E}_{inc}(\theta,\phi)$ as
\begin{equation}
\label{fieldinout}
\vec{E}_{res}(\theta,\phi)=A\vec{E}_{inc}(\theta,\phi),
\end{equation}
where the transfer function $A$ is given by 
\begin{widetext}
\begin{eqnarray}
\label{transfermatrix}
A_{1,j}^t &=  R_z(-\phi)R_y(\theta)TR_z(\phi) = \left[
\begin{array}{ccc}
\ \cos \phi  &\ -\sin \phi &\ 0\ \\
\  \sin \phi &\ \cos \phi &\ 0\ \\
\ 0 &\ 0 &\ 1\ \\
\end{array}
\right ] \left[
\begin{array}{ccc}
\ \cos\theta \ &\ 0 & \ -\sin\theta \\
\ 0\ &\ 1 &\ 0 \\
\ \sin \theta\ &\ 0 &\ \cos\theta\\ \end{array}
\right ] \left[
\begin{array}{ccc}
\ T_p  &\ 0 &\ 0\ \\
\  0 &\ T_s &\ 0 \\
\ 0 &\ 0 &\ T_p\\ 
\end{array}
\right ] \left[
\begin{array}{ccc}
\ \cos \phi  &\ \sin \phi &\ 0\ \\
\  -\sin \phi &\ \cos \phi &\ 0\ \\
\ 0 &\ 0 &\ 1\ \\
\end{array}
\right ] \nonumber \\ 
&= \left[
\begin{array}{ccc}
\ a - b\cos 2\phi  &\ -b\sin 2\phi &\ c\cos\phi\ \\
\  -b\sin 2\phi &\ a + b\cos 2\phi &\ c\sin\phi\ \\
\ -c\cos\phi &\ -c\sin\phi &\ a-b\ \\
\end{array}
\right ].
\end{eqnarray}
\end{widetext} 

For the forward-propagating case, the coefficients of $A$ are given by $a = \frac{1}{2} \left (T_s+T_p \cos\theta \right )$, $b = \frac{1}{2}\left (T_s-T_p \cos\theta \right )$, and $c =  T_p\sin\theta$.  Note that, in general, $\vec{E}_{res}(\theta,\phi)$ would be a superposition of forward and backward propagating waves in the stratified media, though the dominant contribution would come from the forward propagating  waves. In contrast, for the backward propagating waves the coefficients in eq.~\ref{transfermatrix} would be modified with $\theta$ replaced by $\pi - \theta$, and the Fresnel reflection coefficients $R_s$ and $R_p$ being used instead of the transmission ones. Then, the final field can be obtained by integrating Eq.~\ref{fieldinout} over $\theta$ and $\phi$, so that we finally have
\begin{eqnarray}\label{polarint}
\vec{E}(\rho,\psi,z)&=&  i\frac{kfe^{-ikf}}{2\pi}\int_0^{\theta_{max}}\int_0^{2\pi}
\vec{E}_{res}(\theta,\phi)e^{ikz\cos\theta}\nonumber\\& \times & e^{ik\rho\sin\theta\cos(\phi-\psi)}
sin(\theta)\>d\theta d\phi,
\end{eqnarray}
where $r$ is set to $f$ -- the focal length of the lens, and the limit for the $\theta$ integral is set by the numerical aperture of the microscope objective.
The cylindrical coordinate system is chosen for the convenience it offers to track the polarization of the light beam at the output of a high numerical aperture objective, where it is completely modified from the incident polarization \cite{roh05}. For an incident linearly polarized beam of light (polarized along $x$ direction represented by a Jones vector  $\left[\ 1\ 0\ 0\ \right]^T$), the electric field can be written from eq.~\ref{polarint} in  matrix form as 
\begin{eqnarray}\label{fieldout}
\left[
\begin{array}{c}
 {E_x}\\
 {E_y}\\
 {E_z}\\
\end{array}
\right ] &=& C \left[
\begin{array}{lll}
 I_0 + I_2\cos 2\psi  & I_2\sin 2\psi & 2i I_1\cos\psi \\
 I_2\sin 2\psi & I_0 - I_2\cos 2\psi & 2iI_1\sin\psi \\
 -2iI_1\cos\psi & -2iI_1\sin\psi & I_0+I_2 \\
\end{array}
\right ]\nonumber\\  && \times \left[
\begin{array}{c}
 {1}\\
 {0}\\
 {0}\\
\end{array}
\right ]
 =  C
\left[
\begin{array}{c}
 {I_0+I_2\cos 2\psi }\\
 {I_2\sin 2\psi }\\
 {-i2I_1 \cos \psi }\\
\end{array}
\right ].
\end{eqnarray}
Note that this is a general expression that would work for both transmitted and reflected components. More specifically, the values for the transmitted and reflected components of $I_0(\rho),~I_1(\rho)$ and $I_2(\rho)$ would be (suffixes $t$ and $r$ imply transmitted and reflected respectively) given by
\begin{widetext}
\begin{eqnarray}
\label{transmisseqn}
I^t_0(\rho)=\int_0\limits^{min(\theta_{max},\theta_c)}E_{inc}(\theta)
\sqrt{\cos\theta}(T^{(1,j)}_s+T^{(1,j)}_p\cos\theta_j)J_0(k_1\rho\sin\theta)e^{ik_jz\cos\theta_j}sin(\theta)\>d\theta, \nonumber
\end{eqnarray}
\begin{eqnarray}
I^t_1(\rho)=\int_0\limits^{min(\theta_{max},\theta_c)}E_{inc}(\theta)
\sqrt{\cos\theta}T^{(1,j)}_p\sin\theta_j
J_1(k_1\rho\sin\theta)e^{ik_jz\cos\theta_j}\sin\theta\>d\theta, \nonumber
\end{eqnarray}
\begin{eqnarray}
I^t_2(\rho)=\int_0\limits^{min(\theta_{max},\theta_c)}E_{inc}
(\theta)\sqrt{\cos\theta}(T^{(1,j)}_s-T^{(1,j)}_p\cos\theta_j) J_2(k_1\rho\sin\theta)e^{ik_jz\cos\theta_j}\sin\theta\>d\theta,
\end{eqnarray}
\end{widetext}
and 
\begin{widetext}
\begin{eqnarray}
I^r_0(\rho)=\int_0\limits^{min(\theta_{max},\theta_c)}E_{inc}(\theta)\sqrt{\cos\theta}
(R^{(1,j)}_s-R^{(1,j)}_p\cos\theta_j)J_0(k_1\rho\sin\theta)e^{-ik_jz\cos\theta_j}\sin\theta\>d\theta, \nonumber
\label{reflecteqn}
\end{eqnarray}
\begin{eqnarray}
I^r_1(\rho)=\int_0\limits^{min(\theta_{max},\theta_c)}E_{inc}(\theta)\sqrt{\cos\theta}
R^{(1,j)}_p\sin\theta_k J_1(k_1\rho\sin\theta)e^{-ik_jz\cos\theta_j}\sin\theta_1\>d\theta, \nonumber
\end{eqnarray}
\begin{eqnarray}
I^r_2(\rho)=\int_0^{min(\theta_{max},\theta_c)}E_{inc}(\theta)\sqrt{\cos\theta}
(R^{(1,j)}_s+R^{(1,j)}_p\cos\theta_j) J_2(k_1\rho\sin\theta)e^{-ik_jz \cos\theta_j}\sin\theta\>d\theta,
\end{eqnarray}
\end{widetext}
where the $\phi$ integrals have been carried out and are related to Bessel functions $J_n$.
\section{\label{160umprop}Study of radial intensity distribution in a conventional optical trap with a single RI interface}
In this section, we study the variation of radial intensity distribution in a conventional optical tweezers system having a single RI interface in the forward direction of propagation of light. This occurs since most cover slips used in optical trapping are generally refractive index (RI) matched with the microscope immersion oil (RI 1.515), and have thickness between 130 -- 160 $\mu$m so as to obtain high axial trapping depth and also to reduce the effects of spherical aberrations inside the sample. 

\begin{figure}[!h]
 \centering{\includegraphics[scale=0.4]{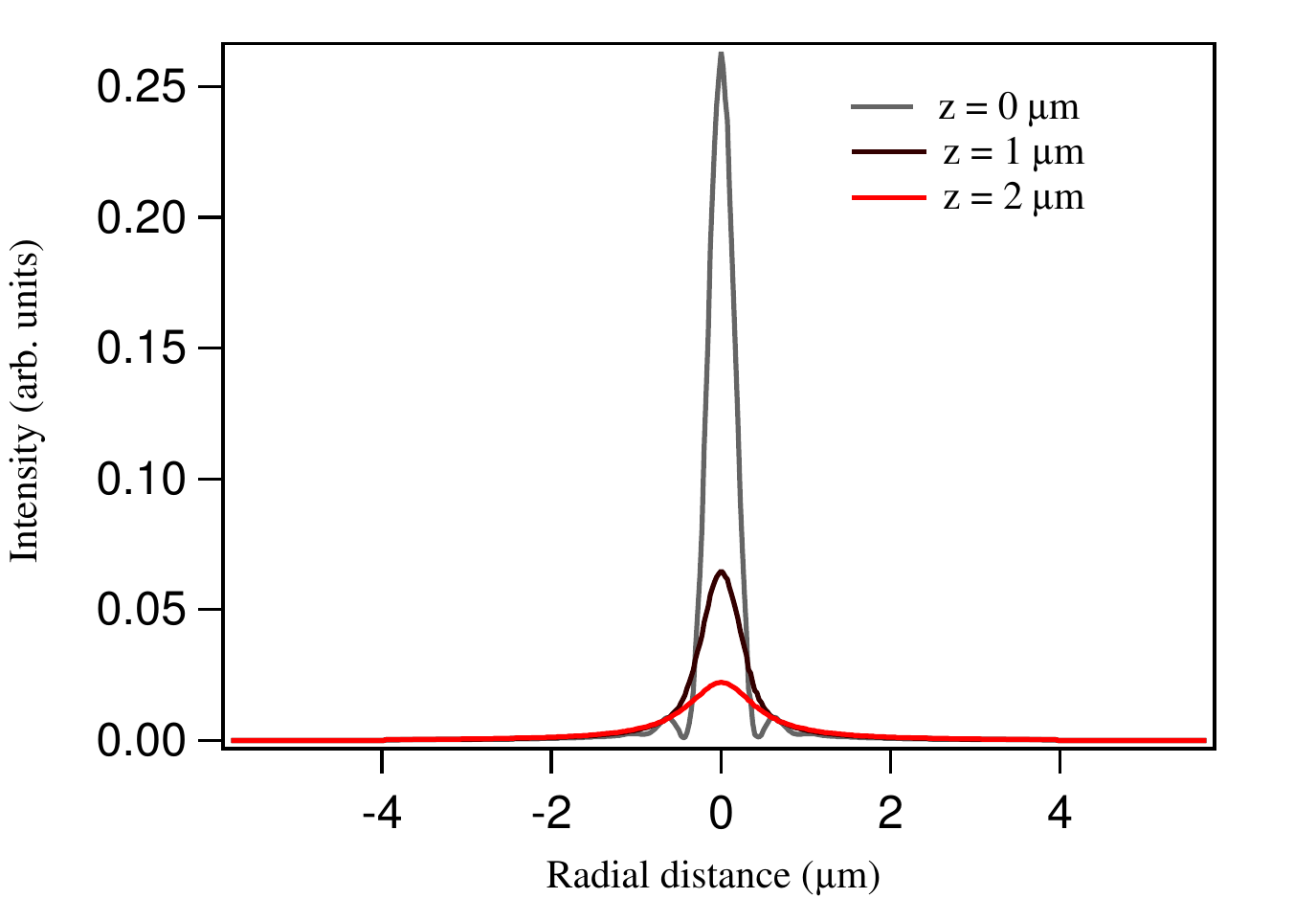}}
\caption[i0i2comp]{(Color online) Plot of  the total intensity from Eq.~\ref{intensitylinpolar} as a function of beam propagation in the $z$-direction for a single RI interface at (a) beam focus, and (b) axial distance of 1 $\mu$m away from focus, (c) axial distance of 2 $\mu$m away from focus.}
\label{intensityplot}
\end{figure}
The total intensity distribution $I(\rho)$ for input $x$-polarized light, considering a superposition of both transmitted and reflected components, is 
given by \begin{equation}
I(\rho) = \ \left|I_0\right|{^2} + \left|I_2\right|{^2}  \pm  2 {\bf Re}(I_0I_2^{\star})\cos 2\psi + 2 \left|I_1\right|^2 (1 \pm \cos 2 \psi)
\label{intensitylinpolar}
\end{equation}
Fig.~\ref{intensityplot} shows the radial ($x-y$) variation of the total intensity as given by Eq.~\ref{intensitylinpolar} at different axial ($z$) distances inside the sample chamber for standard cover slips. The sample thickness in the sample chamber is 20 $\mu$m, and the geometrical beam focus is at an axial  distance of 13 $\mu$m inside the chamber. As the off-axis values of the linear diattenuation term $\mathfrak{D}(\rho) = {(\bf Re}\ (I_0I_2^{\star})+ \left|I_1\right|^2)$ is weak in such cases as shown in Fig.~1 in the main manuscript, the intensity is confined almost entirely in the center. It is observed that the focal spot shows the typical Gaussian structure with weak Airy lobes in the sides, while at $z=1$ and $2~\mu$m, the beam diverges in a manner typical of TEM$_{00}$ Gaussian beams, and one observes a reduction in  height of the central lobe accompanied by an increase of the FWHM of the lobe with increasing axial distance. The Airy lobes are also smeared out when one moves away from the focus, making this a similar to the classic instance of Gaussian beam propagation in air after focusing by a lens (also a single glass-air interface). Particles are thus trapped only in the center, and since the variation of intensity as a function of input polarization is also rather weak, it merely leads to a slight elongation of the focal spot as has been mentioned in literature \cite{roh05}, and does not therefore lead to transportation of particles. Indeed we typically observe clumping of particles in the central region of the beam with time as the sample solution is exposed to the trapping laser \cite{hal12}. Note that the value of $\mathfrak{D}$ also increases with axial distance leading to higher spread of the focal spot. This is the phenomenon that is referred to as spherical aberration which leads to the weakening of axial trapping at large axial distances inside the sample chamber. It is also trivial to note that increasing the thickness of cover slips would not change the radial intensity distribution in this case since there is a single RI interface that is being encountered by the forward propagating light.  
\section{\label{160umprop}Study of axial intensity distribution inside sample chamber for our experimental system}
\begin{figure}[h!t!]
  \centering
 \centering{\includegraphics[scale=0.4]{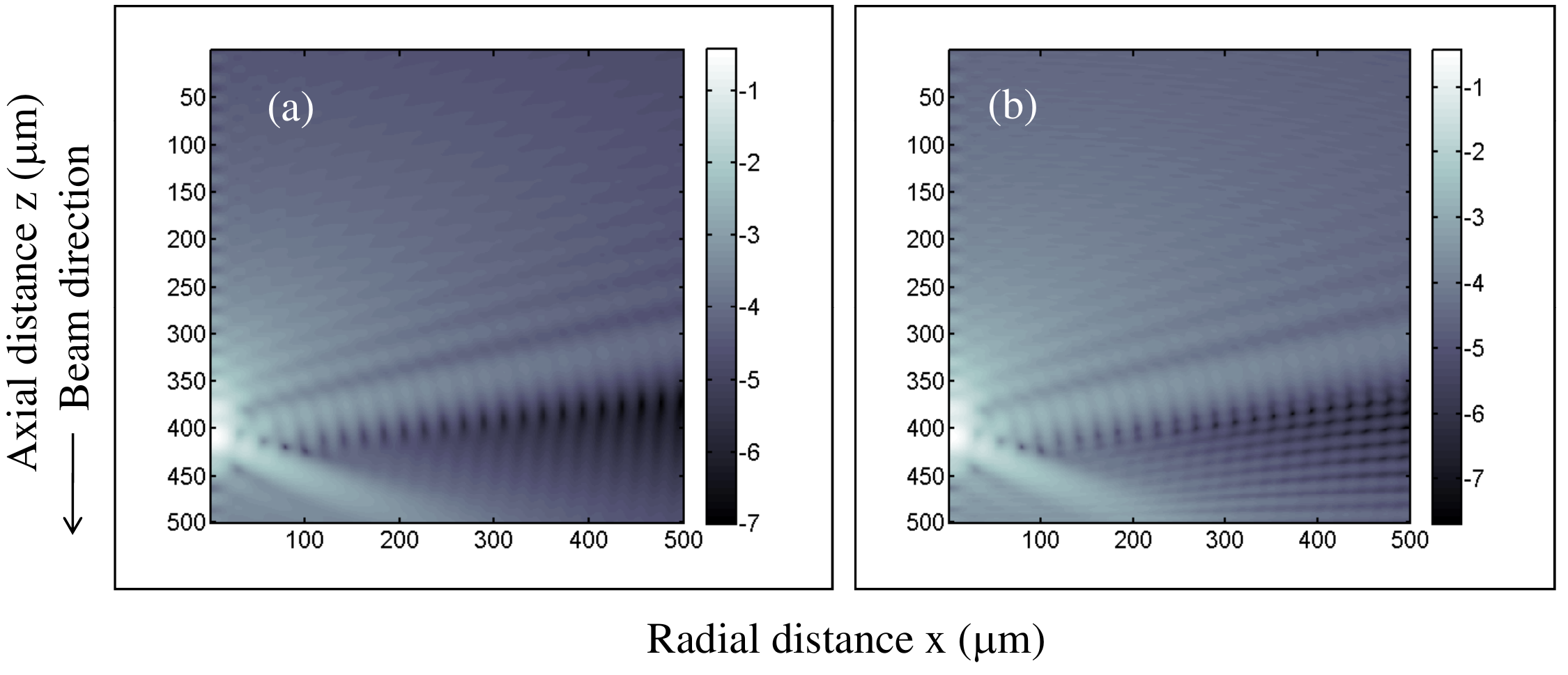}}
  \caption{(a) $xz$ profile of the field inside the sample chamber without a top slide. (b) Axial fringes produced due to reflections from top slide ($0<x<10~\mu$m,$0<z<26~\mu$m). The intensity color bar on the right axes is given in logarithmic scale.}
  \label{axialdist1}
\end{figure}

An additional feature of the electric field inside the sample chamber is the formation of axial fringes due to back-reflected waves from the top slide of our sample chamber. This is shown in the $xz$ plot in Fig.~\ref{axialdist1}. It can be seen that no fringes are seen in the absence of a top slide (Fig.~\ref{axialdist1}a). With a top slide, the location and separation of axial fringes depends on the thickness of the water layer and also on the position of the focus with respect to the top glass slide - water interface. The figures were generated with focal spot 13 $\mu$m inside the sample chamber. Note that the direction of the beam is reversed in these figures compared to ones reported in our manuscript. Off-axis trapping occurs only in the presence of a top slide in the sample chamber where single particles are trapped at the intersection of a radial maxima and an axial maxima. Also, axial fringes are located near the top slide only and die away quickly as one goes farther into the sample solution. This is intuitively understandable considering the fact that we are working with very fast diverging Gaussian beams in this case, and a constructive superposition could be achieved only when the incident beam and reflective surface are very close (up to within 4-5 $\mu$m). 
\section{\label{250umthick}Study of radial intensity distribution inside sample chamber for different cover slip thicknesses}
\begin{figure}[h!t!]
 \centering{\includegraphics[scale=0.35]{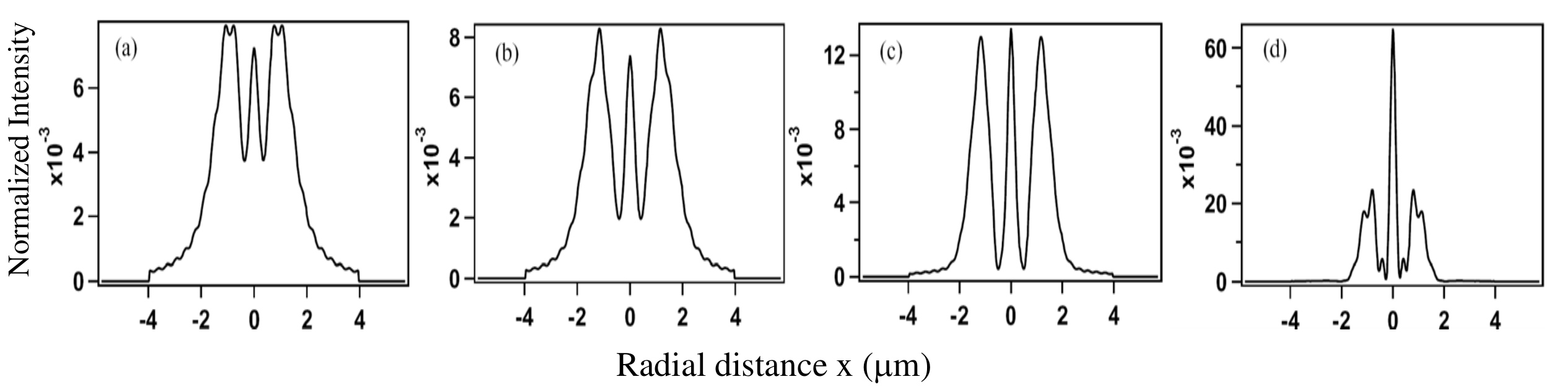}}
\caption[]{(Color online) Simulation of the radial variation of total intensity at axial distance 1 $\mu$m away from the focus for cover slips (RI 1.575) having thickness (a) 160, (b) 200, (c) 250, and (d) 300 $\mu$m.}
\label{intensitythick}
\end{figure}
Since we obtain these effects with a stratified media, an interesting exercise is to study the effect of variation of the thickness of one of the layers of the stratified media - viz.~that of the cover slip - in the radial intensity pattern. This is depicted in Fig.~\ref{intensitythick}. The four sub-plots in the figure are for cover slip thicknesses 160, 200, 250, and 300 $\mu$m respectively (all having RI 1.575) and the radial intensity cross-section are taken at $x-y$ planes at an axial distance of 1 $\mu$m from the beam focus in each case. It is observed that with increase in cover slip thickness, the intensity in the side lobes is higher, with the intensity for 300 $\mu$m thick cover slips being around three times higher than that for 160 $\mu$m thick cover slips. This signifies that particles could be trapped in the side lobes at higher axial distances, i.e. an increase in axial depth in off-axis trapping with increase in thickness. This should thus lead to larger separation of side lobes where stable off-axial trapping would be achieved. Thus, larger distances for particle transportation could be possible using thicker cover slips. However, the limit to the thickness would be set by the maximum allowable focal depth of the trapping objective. It can also be expected that increasing the RI contrast of the stratified media would lead to very similar effects. However, these effects are being investigated presently.

%

